\documentclass{aa}
\usepackage{bibtex/natbib}
\usepackage{xspace}
\usepackage{amssymb}
\usepackage{amsmath}
\usepackage{graphicx}
\usepackage{verbatim} 
\usepackage{fancyhdr}

\def\pP{\textit{`pP'}}

\begin{document}

\title{The outcome of protoplanetary dust growth:\\
pebbles, boulders, or planetesimals?}
\subtitle{III. Sedimentation driven coagulation inside the snow-line}
\authorrunning{A. Zsom et al.}
\titlerunning{Sedimentation driven coagulation}
\author{A. Zsom$^{1}$, C.W. Ormel$^{1}$, C. P. Dullemond$^{1,2}$, Th. Henning$^{1}$}
\institute{Max-Planck-Institute f\"ur Astronomie, K\"onigstuhl 17, D-69117 Heidelberg, Germany. Email: \texttt{zsom@mpia.de} 
\and
Institut f\"ur Theoretische Astrophysik, Universit\"at Heidelberg, Heidelberg, Germany}
\date{\today}

 \abstract
   {The evolution of dust particles in protoplanetary disks determines many observable and structural properties of the disk such as the spectral energy distribution (SED), the appearance of disks, temperature profile, and chemistry. Dust coagulation is also the first step towards planet formation.}
   {We investigate dust growth due to settling in a 1D vertical column of a disk. We gradually build up the complexity of the models by considering the effects of porosity, different collision models, turbulence, and different gas models respectively. This way we can distinguish the effects of these physical processes on particle growth and motion. It is known from the 10 micron feature in disk SEDs, that small micron-sized grains are present at the disk atmosphere throughout the lifetime of the disk. We hope to explain such questions as what process can keep the disk atmospheres dusty for the lifetime of the disk and how does the particle properties change as a function of height above the midplane.}
   {We use a Monte Carlo code to follow the mass and porosity evolution of the particles in time. The used collision model is based on laboratory experiments performed on dust aggregates. As the experiments cannot cover all possible collision scenarios, the largest uncertainty of our model is the necessary extrapolations we had to perform. We simultaneously solve for the particle growth and motion. Particles can move vertically due to settling and turbulent mixing. We assume that the vertical profile of the gas density is fixed in time and only the solid component evolves.}
  {We find that the used collision model strongly influences the masses and sizes of the particles. The laboratory experiment based collision model greatly reduces the particle sizes compared to models that assume sticking at all collision velocities. We find that a turbulence parameter of $\alpha = 10^{-2}$ is needed to keep the dust atmospheres dusty, but such strong turbulence can produce only small particles at the midplane which is not favorable for planetesimal formation models. We also see that the particles are larger at the midplane and smaller at the upper layers of the disk. At 3-4 pressure scale heights micron-sized particles are produced. These particle sizes are needed to explain the 10 micron feature of disk SEDs. Turbulence may therefore help to keep small dust particles in the disk atmosphere.}
 {}

\maketitle

\section{Introduction}
The coagulation of dust particles is the first step towards planet formation. As dust particles are the main source of opacity \citep{Semenov2003}, they also determine many observable quantities of disks, such as the spectral energy distribution and scattered light images (see below). Due to the vertical component of the stellar gravity, dust particles sediment towards the midplane of the disk. The relative settling velocity between particles of different aerodynamical properties drives coagulation and this process was investigated by several authors \citep{Safronov1969, Nakagawa1986, Schrapler2004, Dullemond:2004p325, Dullemond2005}.

Observational evidence of the vertical sedimentation of grains exists for a large number of disks, although such evidence is usually indirect \citep{Henning2011}. Sub-micron grains are present at the disk surfaces as shown by scattered light images in the optical and near infrared (NIR) wavelengths. Such multi-wavelength scattered light images provide evidence for grain growth (\cite{Watson2007} and references therein). \cite{Pinte2007} showed by reproducing multi-band images of the binary system of GGTau that the dust scale height for 10 micron-sized particles is roughly half of that for micron-sized particles. The spectral energy distribution (SED) is also affected by settling. \cite{D'Alessio2006} showed that in order to explain the median SEDs of classical TTauri stars, the dust to gas ratio has to be reduced by a factor of 10 at the disk atmosphere compared to the standard value. There are also indications that the settling of grains is correlated with the age of the disk \citep{Sicilia-Aguilar2007}. However, the connection between the exact shape of the 10 micron feature of SEDs and sedimentation is not well understood \citep{Dullemond2008, Juh'asz2010}. \cite{Apai2005} showed evidence for settling in disks around brown dwarf stars. They concluded that growth, crystallization and settling of dust happens around low mass stars in a similar manner as around intermediate and solar mass stars. This suggests that the first stage of planet formation is a robust process occurring in all kinds of disks.

Sedimentation also affects the vertical temperature structure of the disk. The simulations of \cite{Aikawa2006} showed that as the dust particles sediment towards the midplane, the opacity is reduced, and the temperature of the gas decreases. As the stellar radiation can now penetrate deeper in the disk, the temperature at intermediate heights increases. The change in the density and temperature structure as well as the radiation field naturally influences the chemistry of the disk atmospheres \citep{Bergin2007}. 

Recently \cite{Vasyunin2011} investigated the effects of dust growth on disk chemistry and found that the main effect of coagulation on the disk chemical composition comes mostly from sedimentation in their models.

Dust sedimentation not only affects the upper layers of the disk, but also the midplane regions. As long as the dust to gas ratio is much smaller than unity, the dust can be treated as a passive tracer in the gas flow. However, if the dust to gas ratio becomes comparable to unity, the dust will influence the gas. The dust can be accumulated by sedimentation around the midplane of the disk. In such a scenario, a shear is present between the dusty midplane layer and the less dusty upper layers. This shear triggers Kelvin-Helmholtz instability which develops into turbulence. This process was first recognized by \cite{Weidenschilling1980} and is still under active investigation (e.g. by \cite{Johansen2006}, \cite{Chiang2008}, \cite{Lee2010}).

A collaboration started between the lab-community and the modelers to better constrain the dust evolution in protoplanetary disks, using a realistic collision model that is based on the laboratory experiments. In \cite{Guttler2010} (henceforth Paper I), we introduced this collision model. In \cite{Zsom2010} (henceforth Paper II), we used this collision model for the first time in the Monte Carlo (MC) method of \cite{Zsom2008} (henceforth ZsD08). The models of Paper II were local box models, meaning that the dust evolution was only followed at one location of the disk. These models showed that bouncing plays an important role in dust evolution. 

We further develop these models to simulate a 1D vertical column in the disk, thus investigating sedimentation-driven coagulation. We want to better understand the process of sedimentation and the role of particular physical phenomena like porosity of the aggregates, collision models and turbulence. 

Previous work by \cite{Dullemond2005} (henceforth DD05) showed that without a mechanism that reduces the sticking probability of particles in the upper layers of the disk, or without a continuous source of small particles, the observed spectral energy distributions (SED) of TTauri stars should exhibit a very weak IR excess. In contrast, the observed SEDs of TTauri stars have strong IR excesses (e.g. \cite{Furlan2005, Kessler-Silacci2006, Bouwman2008}). Therefore some grain-retention mechanism is needed to explain the SEDs. 

Previous models of grain evolution assumed a continuous cycle of growth and fragmentation, which provides the necessary amount of small particles (e.g. \cite{Brauer2008a}, \cite{Birnstiel2009}). However, Paper I and II showed that particle evolution is halted by bouncing and no cycle of growth and fragmentation is present. In this paper we simulate dust evolution driven by Brownian motion, turbulence, and sedimentation in a 1D vertical column of the inner disk and the additional physics of Paper I and II are included. We investigate the time evolution of sedimentation-driven coagulation, and search for ways that can keep a sufficient amount of the small dust particles levitated at several pressure scale heights to explain the observed SEDs of young stars. 

The paper is organized as follows. In Sect. \ref{sec:num} we describe the numerical method used to follow the particle motion and coagulation. We validate the code and increase the complexity of the model step-by-step in Sect. \ref{sec:intro}. We show the results in Sect. \ref{sec:res}, finally we discuss those results in Sect. \ref{sec:disc} and provide a summary in Sect. \ref{sec:concl}.
 
\section{Numerical method}
\label{sec:num}
\subsection{Basic considerations}
\label{subsec:bas}
The local box approach (or ``particle-in-a-box'' approach) in Paper II is based on two assumptions. 1.) The particles are homogeneously mixed inside the box. 2.) Particles do not enter or leave the box, i.e. it is closed. Due to these two assumptions, it was not necessary to follow the exact location of the particles. 

In the models considered here, however, we place such boxes (or grid cells) on top of each other to simulate a 1D column in the disk and follow how particles settle towards the midplane. Inevitably, particles move from box to box during this process. Therefore the assumption that particles cannot enter or leave the boxes has to be relaxed. The first assumption of the method in Paper II is kept, we still assume that during the coagulation calculation, the particles inside a given box are homogeneously distributed (however, for particle motions, the individual positions of the particles are used). The second assumption is modified in the following way. 2.) The simulated column is closed, e.g. particles inside the column can move freely vertically.
However neither do \textit{new} particles enter from the ``outside'', nor do particles from inside the column \textit{leave}. As particles move through the boxes, it is necessary to follow the position of the particles (see Sect. \ref{subsec:pos}) as we must find out in which box a particle is located.

The motion of particles imposes a limit on the time step of the simulation. We do not want the particles to move more than one box in a time step. A sedimenting particle should have the possibility to interact with all other particles along its way, it should not skip over boxes thus avoiding the particles in it. Therefore, we use an adaptive time stepping method. The maximum of all particle velocities is obtained ($v_{\mathrm{max}}$), and since we know the height of the boxes ($h_{\mathrm{box}}$), the maximal (safe) time step can be determined as
\begin{equation}
\Delta t = C \frac{h_{\mathrm{box}}}{v_{\mathrm{max}}},
\end{equation}
where $C$ is the Courant number which we typically set to be $0.1$. 

The code schematically performs the following steps:
\begin{enumerate}
\item The velocities of the particles are calculated.
\item A safe time step is determined to avoid particle `jumps'.
\item The position of the particles is updated using their velocities, their previous positions, and the time step.
\item We determine the box in which each particle resides.
\item We call the coagulation subroutine described in Paper II to calculate the evolution of the particles separately in each box for the given time step. There can be multiple collisions per time step.
\end{enumerate}

\subsection{Initial conditions}
\label{subsec:inicond}
We assume that the gas density profile is constant in time during the simulation. This assumption is valid if the simulated time is less or comparable to the viscous timescale of the gas. The viscous timescale can be calculated as
\begin{equation}
t_{\mathrm{vis}}=r^2/ \nu_T,
\end{equation}
where $r$ is the distance from the central star, $\nu_T$ is the turbulent viscosity. The value for $t_{\mathrm{vis}}$ at 1 AU varies between $10^3$ and $10^7$ yr depending on $\nu_T$. We assume that turbulence is parameterized by the \cite{Shakura1973} $\alpha$ parameter
\begin{equation}
\nu_T=\alpha c_s H_g,
\label{eq:nuT}
\end{equation}
where $c_s$ is the isothermal sound speed, and $H_g$ is the pressure scale height of the gas disk. The turbulence parameter $\alpha$ reflects the strength of the turbulence in the disk. Typical values range between $\alpha=10^{-6}$ and $10^{-2}$, where the former corresponds to the turbulent strength in dead zones, the latter describes turbulence in disk atmospheres. In this paper, we assume that $\alpha$ is constant as a function of height, which may be changed in future work (see Sect. \ref{sec:alpha}).

The vertical structure of the disk is determined by the equilibrium between the vertical component of the gravitational force and the vertical pressure gradient in the gas. If the disk mass ($M_{\mathrm{disk}}$) is much smaller than the mass of the star ($M_*$), and the vertical thickness of the disk ($H_g$) is a small fraction of the radial distance (both conditions are safely met for the disk parameters described below), then the vertical density can be approximated as
\begin{equation}
\rho_g (r,z) = \frac{\Sigma(r)}{\sqrt{2 \pi} H_g}\exp(-z^2/2H_g^2),
\end{equation}
where $\Sigma(r)$ is the gas surface density at distance $r$, and $z$ is the height above the midplane. In this paper we choose $M_* = 0.5 M_\odot$, $r=1$ AU, $\Sigma(1$ AU$)=100$ g/cm$^2$ similarly to DD05. The pressure scale height can be calculated as
\begin{equation}
H_g = c_s/\Omega,
\end{equation}
where $\Omega$ is the orbital frequency at 1 AU. The isothermal sound speed is
\begin{equation}
c_s = \sqrt{\frac{k_B T}{\mu m_p}},
\end{equation}
where $k_B$ is the Boltzmann constant, $\mu$ is the molecular weight, which is 2.3 for molecular gas, $m_p$ is the mass of the proton, and $T$ is the temperature of the gas, which is 200 K for the stellar and disk parameters considered above (see DD05). We assume the temperature to be constant as a function of height. This is a reasonable assumption well below the photosphere if the temperature of the gas is solely determined by the stellar irradiation and the thermal coupling to the grains. The height of the photosphere may (and presumably will) change as the particles rain out, but we do not include this effect in this paper.

We need to quantify the proper number of particles and boxes to be used. As we know from ZsD08, a low number of particles is not desirable because the physics of the collision kernel will not be reproduced properly. However, many particles make the code computationally expensive and unfeasible to run. If too few boxes are used, the Gaussian profile of the gas disk will not be resolved properly. If too many boxes are used, sufficient number of particles per box is needed, which again renders the simulation computationally expensive. We found by numerical experiments that using $10^5$ particles and 40 evenly spaced boxes is a good compromise. We simulate 4 pressure scale heights (0.16 AU above the midplane), therefore there are 10 boxes per pressure scale height to resolve the Gaussain profile of the gas. $10^5$ particles are also enough to remove numerical artifacts from the simulation and reproduce the physics of the collision kernel properly. Using $10^5$ particles and 40 boxes also mean that there is at least 1 particle in the uppermost box in the beginning of the simulation.



\subsection{Position update}
\label{subsec:pos}
The vertical position of the particles are determined by vertical settling and turbulent diffusion. In principle, Brownian motion also contributes to the change of particle positions, but its effect is negligible compared to the other two effects.

The equation governing the diffusion and settling of the dust in a non-homogenous gas density field is \citep{Dubrulle1995, Fromang:2006p324, Ciesla2010}
\begin{equation}
\frac{\partial \rho_d}{\partial t} = \frac{\partial}{\partial z} \left[D_d \rho_g \frac{\partial}{\partial z}
\left( \frac{\rho_d}{\rho_g} \right)\right] + \frac{\partial}{\partial z} (\Omega^2 t_s \rho_d z),
\end{equation}
or equivalently:
\begin{equation}
\frac{\partial \rho_d}{\partial t} = \frac{\partial}{\partial z}\left( D_d \frac{\partial \rho_d}{\partial z}\right) - \frac{\partial}{\partial z}\left( \rho_d \times D_d \frac{1}{\rho_g}\frac{\partial\rho_g}{\partial z}\right) + \frac{\partial}{\partial z} (\rho_d \times z \Omega^2 t_s)
\label{eq:diff}
\end{equation}
where $\rho_d$ is the dust density, $D_d$ is the diffusion coefficient of the dust (see the next paragraph) and $t_s$ is the stopping time of the particle. The stopping time is the timescale a particle needs to react to the changes of the surrounding gas. 
We define the dimensionless Knudsen number as
\begin{equation}
Kn = \frac{a}{\lambda_{\mathrm{mfp}}},
\end{equation}
where $a$ is the size of the aggregate, and $\lambda_{\mathrm{mfp}}$ is the mean free path of the gas. A particle is in the Epstein regime if $Kn < 1$ (to be more precise, if $a < \frac{9}{4}\lambda_{\mathrm{mfp}}$), where the stopping time is (\cite{Epstein1924}):
\begin{equation}
t_{s} = t_{\mathrm{Ep}} = \frac{3 m}{4 v_{\mathrm{th}} \rho_g A},
\label{eq:ts1}
\end{equation}
where $m$ and $A$ are the mass and the aerodynamical cross-section of the particle, and $v_{\mathrm{th}}$ is the thermal velocity. If the Knudsen number is greater than 1 (at high gas densities where the mean free path is low or in the case of large particles), the Stokes regime applies and the stopping time becomes
\begin{equation}
t_s = t_{\mathrm{St}} = \frac{3 m}{4 v_{\mathrm{th}} \rho_g A} \times \frac{4}{9} \frac{a}{\lambda_{\mathrm{mfp}}}.
\label{eq:ts2}
\end{equation} 

The first term on the right hand side of Eq. \ref{eq:diff} is the diffusion term. Using only this term, particles with $t_s = 0$ (tracers) would be homogeneously distributed as a function of height over several diffusion timescales. The first and the second term together on the right hand side ensures that the tracer particles will be distributed according to the background gas density field. The third term describes the settling of the particles. Equation \ref{eq:diff} is valid if the motion of the dust does not influence the motion of the gas (the back-reaction from the dust to the gas is negligible). This condition is met if the dust to gas ratio is $\ll 1$.

We note that we do not solve for the dust density directly. We follow the motion of dust \textit{particles}, each of which represents a portion of the total dust mass inside the column. Thus we derive the corresponding velocities (or fluxes) for the first, second, and third terms of Eq. \ref{eq:diff} to calculate the position update of the particles.

We calculate $D_d$, the diffusion coefficient of the dust, and define the diffusion velocity, $v_{D1}$. The diffusion coefficient of the gas can be defined as \citep{Shakura1973}
\begin{equation}
D_g =\nu_T =\alpha c_s H_g.
\end{equation}
Based on \cite{Youdin:2007p576}, the diffusion coefficient of the dust can be calculated as
\begin{equation}
D_d = D_g/(1+St^2),
\end{equation}
where $St$ is the Stokes number 
\begin{equation}
St = t_s \Omega.
\end{equation}
The average displacement of a particle in 1D during the time step of $\Delta t$ then is
\begin{equation}
L = \sqrt{2 D_d \Delta t}.
\end{equation}
The real displacement of the particle ($\Delta z$) is drawn from a Gaussian distribution which has zero mean and a half width of $L$. The ``diffusion velocity'' can then be calculated as
\begin{equation}
v_{D1} = \Delta z/\Delta t.
\end{equation}
This velocity component tries to smear out dust concentrations. It is important to note that the real, physical velocity of the particle during turbulent diffusion changes randomly every time the aggregate interacts with a turbulent eddy. The diffusion velocity defined above is a numerical construct to calculate the time-averaged velocity of the particle during a time-interval $\Delta t$.

The second term on the right side of Eq. \ref{eq:diff} results in a systematic velocity term which pushes particles towards the density maxima of the gas. This velocity can be determined by
\begin{equation}
v_{D2} = D_d\frac{1}{\rho_g}\frac{\partial\rho_g}{\partial z}. 
\end{equation}
Using these two velocity components ($v_{D1}$ and $v_{D2}$), the particles will be distributed according to the gas density profile in a diffusion timescale.

The fact that particles with $t_s>0$ settle towards the midplane and have a scale height less than $H_g$ is the result of the third term of Eq. \ref{eq:diff}, the settling velocity. The settling velocity of a particle is given by
\begin{equation}
v_{\mathrm{set}}=-z \Omega^2 t_s.
\label{eq:set}
\end{equation}

The new position of the particles can then be determined by using these three velocity terms:
\begin{equation}
z = z_{\mathrm{old}} + (v_{D1}+v_{D2}+v_{\mathrm{set}})\Delta t.
\end{equation}
This description os identical to the one used in \cite{Ciesla2010}.

\begin{table*}
\caption{Overview and results of all the sedimentation simulations.}
\label{table:sedi}      
\begin{center}   
\small
\begin{tabular}{l c c c c c c c c c c}        
\hline\hline                 
ID		&$\Sigma_{g}$	&Coll. model	&Por. model	&$\alpha$&$\Psi_{\mathrm{max}}$	&$<\Psi_{\mathrm{fin}}>$ &$<m_{\mathrm{fin}}>$&$<St>$&$H_p$  \\ 
		&[g/cm$^2$]	&			&			&		&					&					&[g]				   &		 &[$H_g$]\\
(1)		&(2)			&(3)			&(4)			&(5)		&(6)					&(7)					&(8)				   &(9)	 &(10)	\\
\hline                        
DD		&100		&hit\&stick only	&compact		&0		&1				&1 					&$2.5\times 10^{-3}$		&$3.5\times 10^{-3}$	&0\\
DDa		&100		&hit\&stick only	&Okuzumi	&0		&$10^7$			&$1.1\times10^6$ 	&$1.8\times 10^{8}$		&$1.6$				&0\\
SB1		&100		&simpl. Br.		&Okuzumi	&0		&$10^4$			&$2\times10^3$ 	&$2.2\times 10^{-3}$		&$8.0\times 10^{-5}$		&0\\
SB2		&100		&simpl. Br.		&Okuzumi	&$10^{-6}$&$10^4$			&$1.6\times10^3$ 	&$3.2\times 10^{-3}$		&$7.0\times 10^{-5}$		&0.2\\
SB3		&100		&simpl. Br.		&Okuzumi	&$10^{-4}$&$5\times10^3$	&21 				&$2.2\times 10^{-4}$		&$5.0\times 10^{-4}$		&0.6\\
CB1		&100		&compl. Br.		&Okuzumi	&$10^{-4}$&$10^3$			&3 				&$1.2\times10^{-2}$		&$4.5\times 10^{-3}$		&0.2\\
CB2		&100		&compl. Br.		&Ormel		&$10^{-4}$&$10$			&2 				&$1.2\times10^{-2}$		&$4.7\times 10^{-3}$		&0.2\\
CB3		&100		&compl. Br.		&Okuzumi	&$10^{-2}$&$3\times10^2$	&3				&$3.6\times 10^{-7}$		&$1.6\times 10^{-4}$		&0.95\\
CB4		&100		&compl. Br.		&Ormel		&$10^{-2}$&$10$			&2				&$8.9\times 10^{-8}$		&$1.2\times 10^{-4}$		&0.95\\
\hline
LD1		&10		&compl. Br.		&Okuzumi	&$10^{-4}$&$5\times10^2$	&2				&$8.9\times 10^{-4}$		&$9.7\times 10^{-3}$		&0.2\\
LD2		&10		&compl. Br.		&Okuzumi	&$10^{-2}$&$5\times10^1$	&23				&$2.3\times10^{-9}$		&$1.3\times10^{-4}$		&0.95\\
\hline
HD1		&1000	&compl. Br.		&Okuzumi	&$10^{-4}$&$5\times10^4$	&6				&$8.3\times10^{-2}$		&$6.7\times10^{-4}$		&0.5\\
HD2		&1000	&compl. Br.		&Okuzumi	&$10^{-2}$&$10^3$			&3				&$1.2\times 10^{-5}$		&$4.6\times10^{-5}$		&0.95\\
\hline
\end{tabular}
\end{center}
Col. 1 is the ID of the simulations (DD - Dullemond\& Dominik models, SB - simplified Braunschweig model, CB - complete Braunschweig model, LD - low density model, HD - high density model), col. 2 is the gas surface density, col. 3 describes the used collision model (hit\&stick, simplified Braunschweig model, or complete Braunschweig model), col. 4 indicates the used porosity model (based on \cite{Ormel:2007p93} or \cite{Okuzumi2009a}), col. 5 describes the value of the turbulence parameter $\alpha$, col. 6 is the maximum enlargement parameter of the aggregates during the simulation, col. 7, 8, and 8 are the final average enlargement parameter, mass and Stokes number of the particles respectively, and col. 10 is the scale height of the dust expressed in the scale height of the gas. 
\end{table*}
 
\subsection{Coagulation}
\label{subsec:coag}
The collision model used in this work is similar to the one used in Paper II. There are however two differences. The first difference is the additional source of relative velocity due to differential vertical settling (see Eq. \ref{eq:set}):
\begin{equation}
\Delta v_S = | v_{\mathrm{set1}}-v_{\mathrm{set2}} |.
\end{equation}

The second difference concerns the calculation of the aerodynamical cross section which is used to calculate the stopping time. In Paper II we used the geometrical cross section of the particles \citep{Ormel:2007p93}
\begin{equation}
A=r_c^2 \pi \Psi^{2/3},
\label{eq:cross1}
\end{equation}
where $r_c$ is the compact radius of the aggregate (assuming that the mass of the particle is contained in a compact sphere of radius $r_c$), and $\Psi$ is the enlargement parameter. $\Psi=1$ for compact particles and the value is higher for fluffy particles \citep{Ormel:2007p93}. This formula works well for particles with fractal dimension above 2. However, we use the porosity model of \cite{Okuzumi2009a}, and their model produces aggregates with fractal dimension below 2. If one calculates the stopping time of such an aggregate in the Epstein regime (Eq. \ref{eq:ts1}) using the formula in Eq. \ref{eq:cross1}, one gets that the stopping time is less than the stopping time of a monomer. This is clearly unphysical. The reason for this low stopping time (large area) is that Eq. \ref{eq:cross1} does not take into account the empty space between the `fractal branches'. To avoid such unphysical results for aggregates with low fractal dimensions, we use the aerodynamical cross section as defined in Eq. 47 in \cite{Okuzumi2009a}.

\begin{figure}
\centering
  \includegraphics[width=0.49\textwidth]{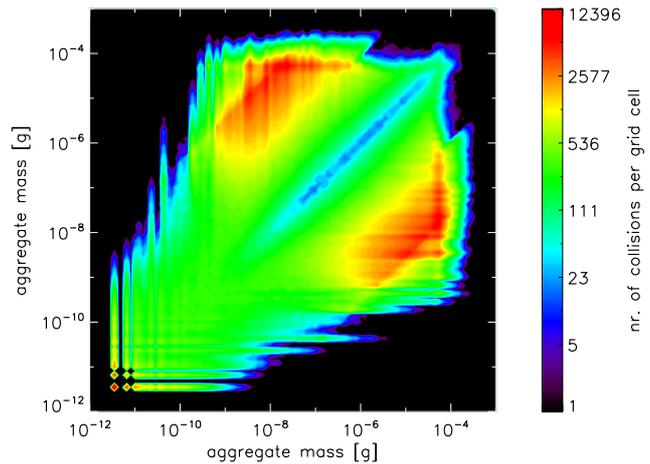}
  \caption{The collision rate between dust aggregates as a function of particle mass to particle mass measured one simulation of Paper II (the MMSN model with ID Mt1d-4m100). Initially particles grow due to Brownian motion and collisions between equal sized aggregates are frequent. When turbulent relative velocity dominates over Brownian motion (masses above $10^{8}$ g), collisions between equal sized aggregates are less frequent and collisions between different sized aggregates are dominant.}
  \label{fig:coll_rate}
\end{figure}

\begin{figure*}
\centering
  \includegraphics[width=0.75\textwidth]{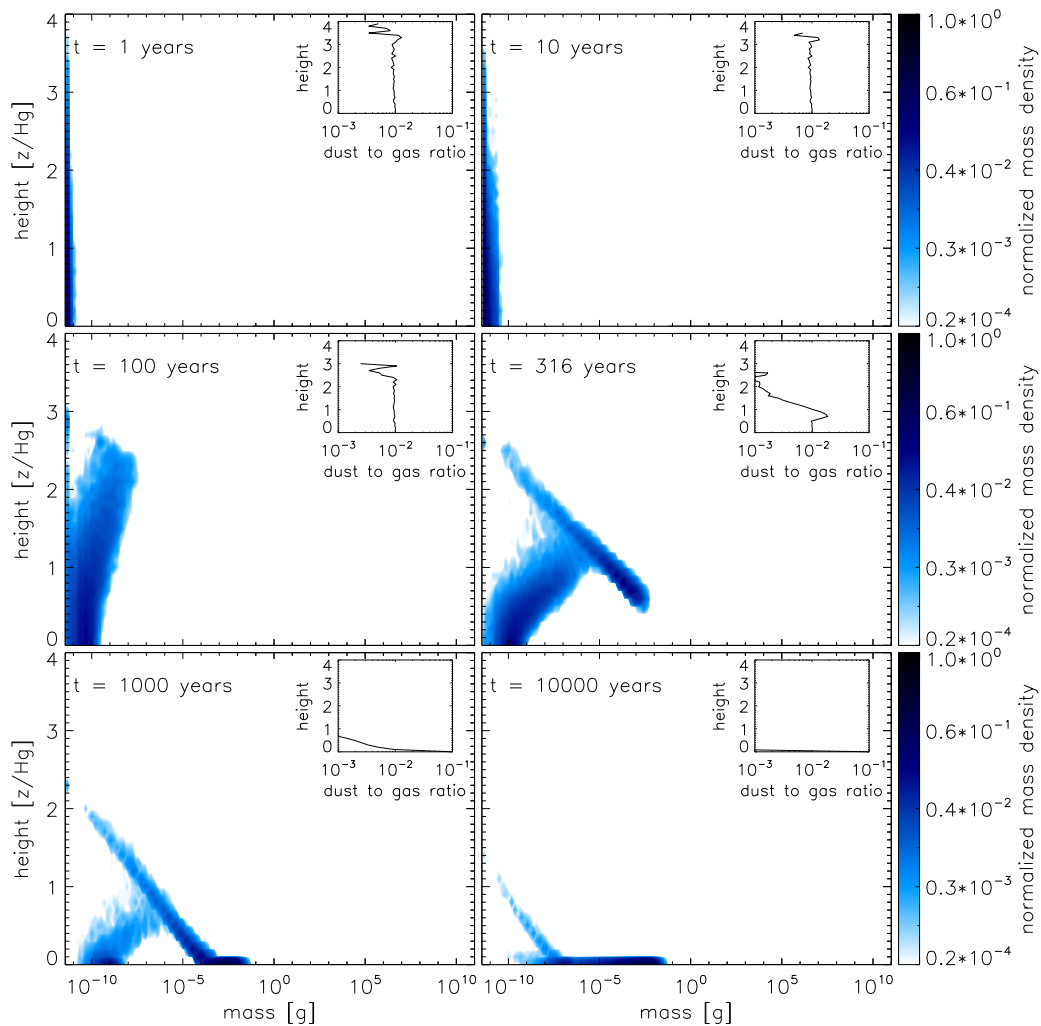}
  \caption{The mass distribution at $t=1$, 10, 100, 316, $10^3$ and $10^4$ yr for the DD model. The x axis is the mass of the aggregates in grams, the y axis is the height above the midplane expressed in units of the pressure scale-height. The contours represent the normalized mass density of the dust. The sub-figures illustrate the dust to gas ratio (x axis) as a function of height above the midplane (y axis)}
  \label{fig:mass_DD05}
\end{figure*}

\begin{figure*}
\centering
  \includegraphics[width=0.75\textwidth]{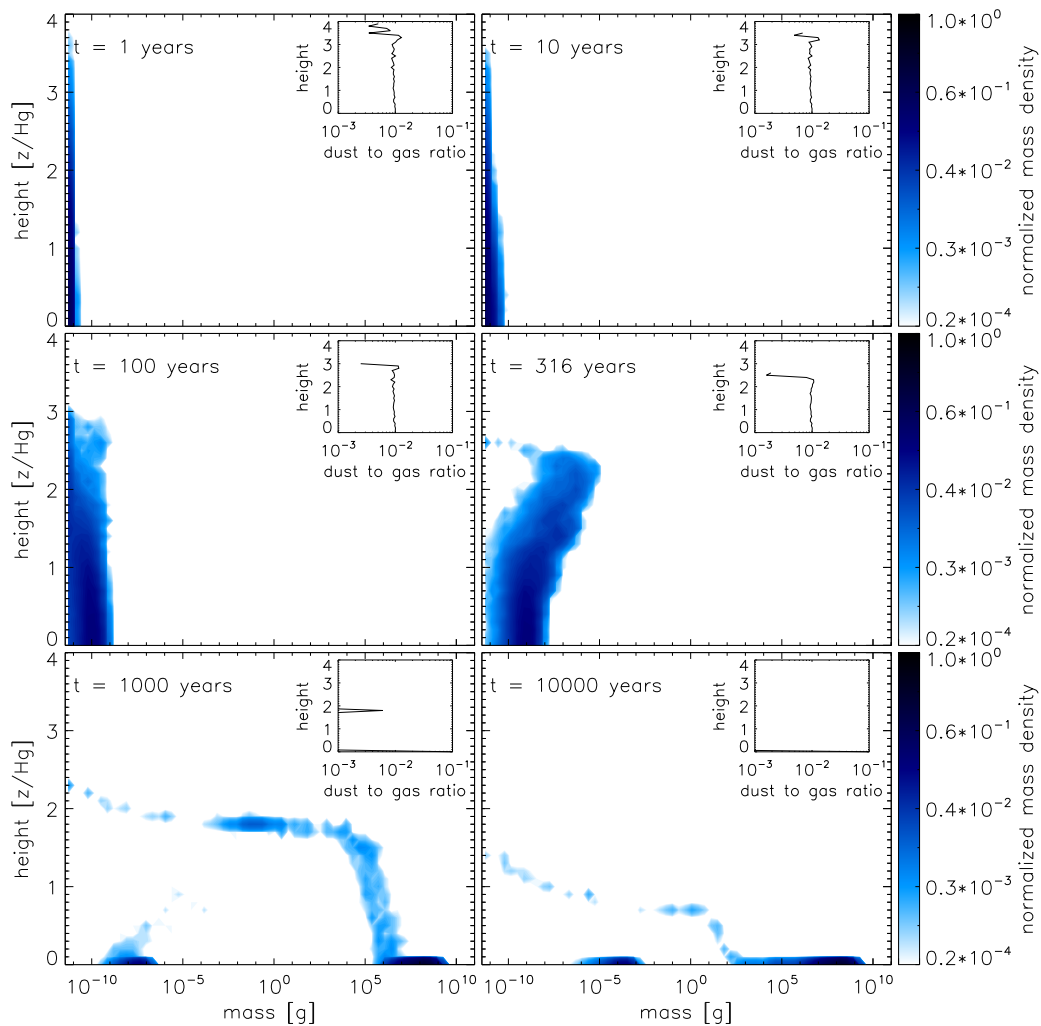}
  \caption{Mass distribution for the DDa model using the Okuzumi hit\&stick porosity model. The axes and contours are the same as in Fig. \ref{fig:mass_DD05}.}
  \label{fig:mass_DD05_ok}
\end{figure*}

\begin{figure*}
\centering
  \includegraphics[width=0.75\textwidth]{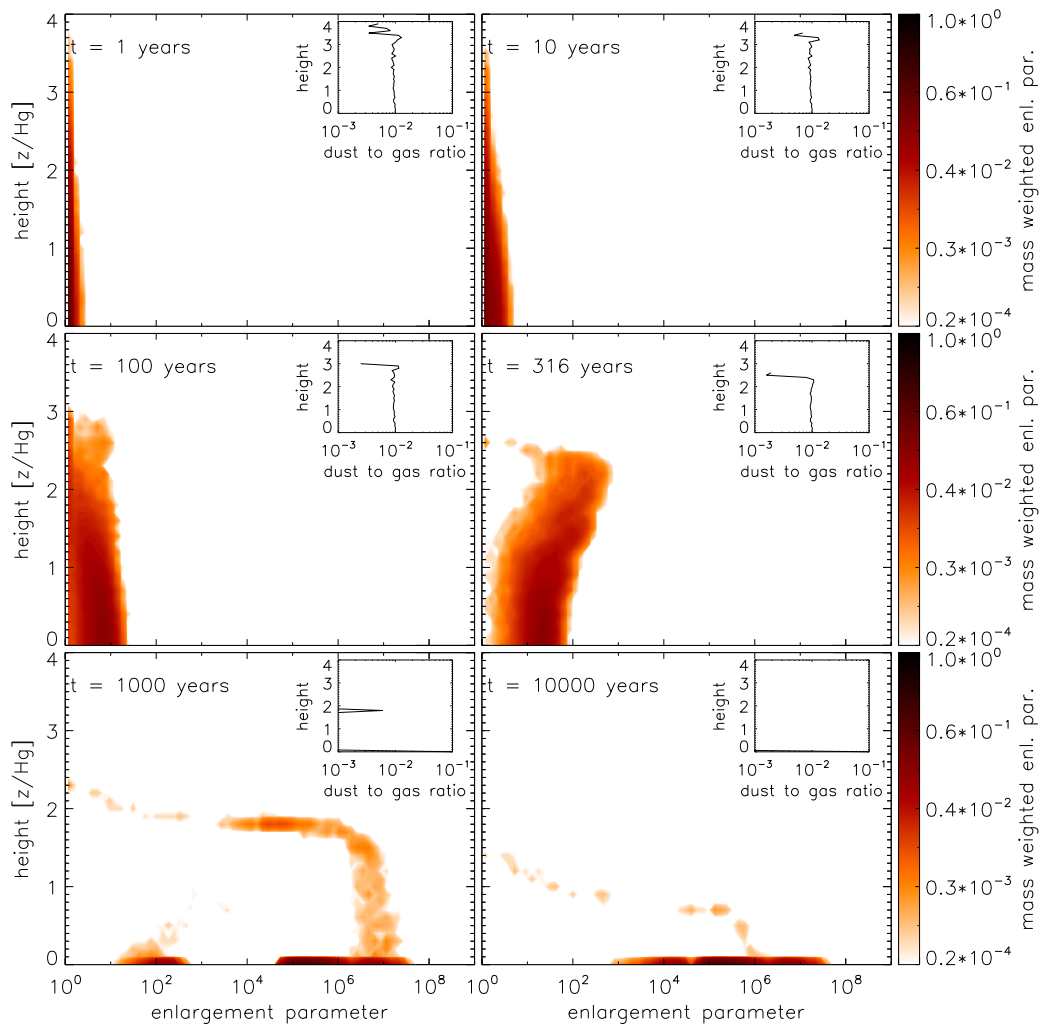}
  \caption{Enlargement parameter distribution for the DDa model using the Okuzumi hit\&stick porosity model. The x axis here represents the enlargement parameter of the aggregates, and the contours show the normalized mass-weighted enlargement parameter of the particles.}
  \label{fig:enpar_DD05_ok}
\end{figure*}

\begin{figure*}
\centering
  \includegraphics[width=0.75\textwidth]{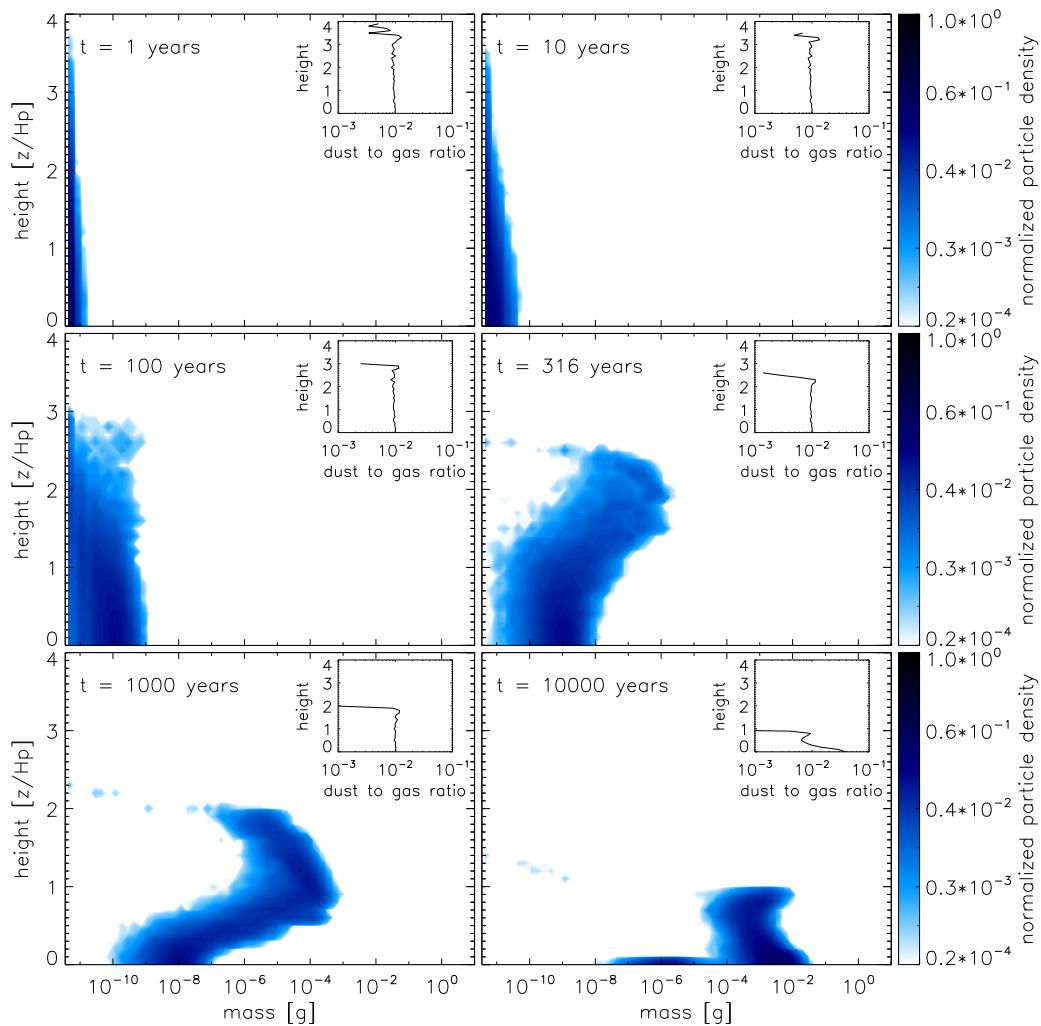}
  \caption{Mass distribution for the simplified Braunschweig collision model using the Okuzumi porosity prescription without turbulent mixing (SB1). The axes and contours are similar to Fig. \ref{fig:mass_DD05}. Notice the x axis ranges from $10^{-12}$ g until 10 g in this figure.}
  \label{fig:mass_simpl_not_or}
\end{figure*}

\begin{figure*}
\centering
  \includegraphics[width=0.75\textwidth]{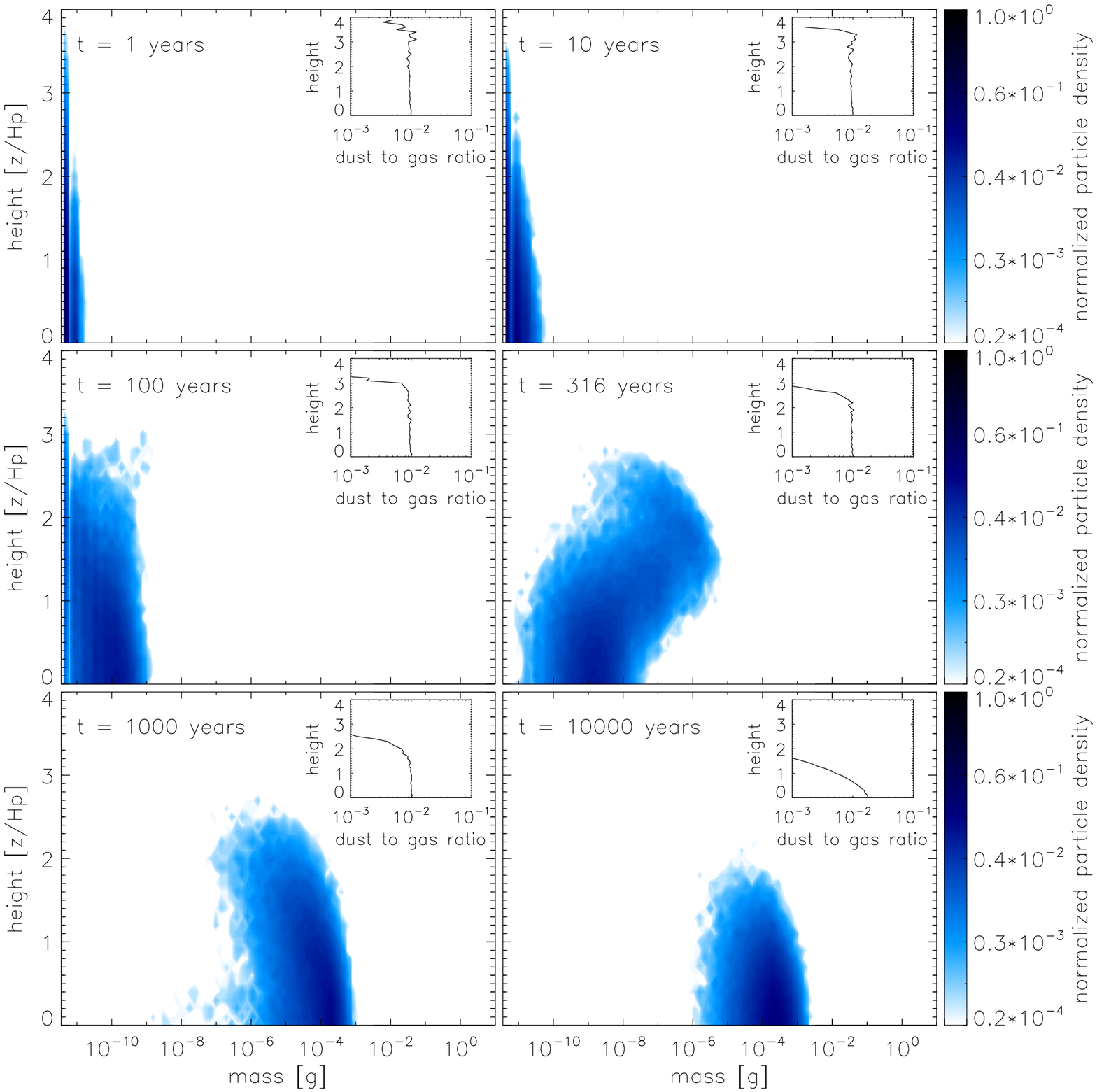}
  \caption{Mass distribution for the simplified Braunschweig model using the Okuzumi porosity prescription with $\alpha=10^{-4}$ (SB3). The axes and contours are the same as in Fig. \ref{fig:mass_simpl_not_or}.}
  \label{fig:mass_simpl_1d-4_or}
\end{figure*}

\begin{figure*}
\centering
  \includegraphics[width=0.75\textwidth]{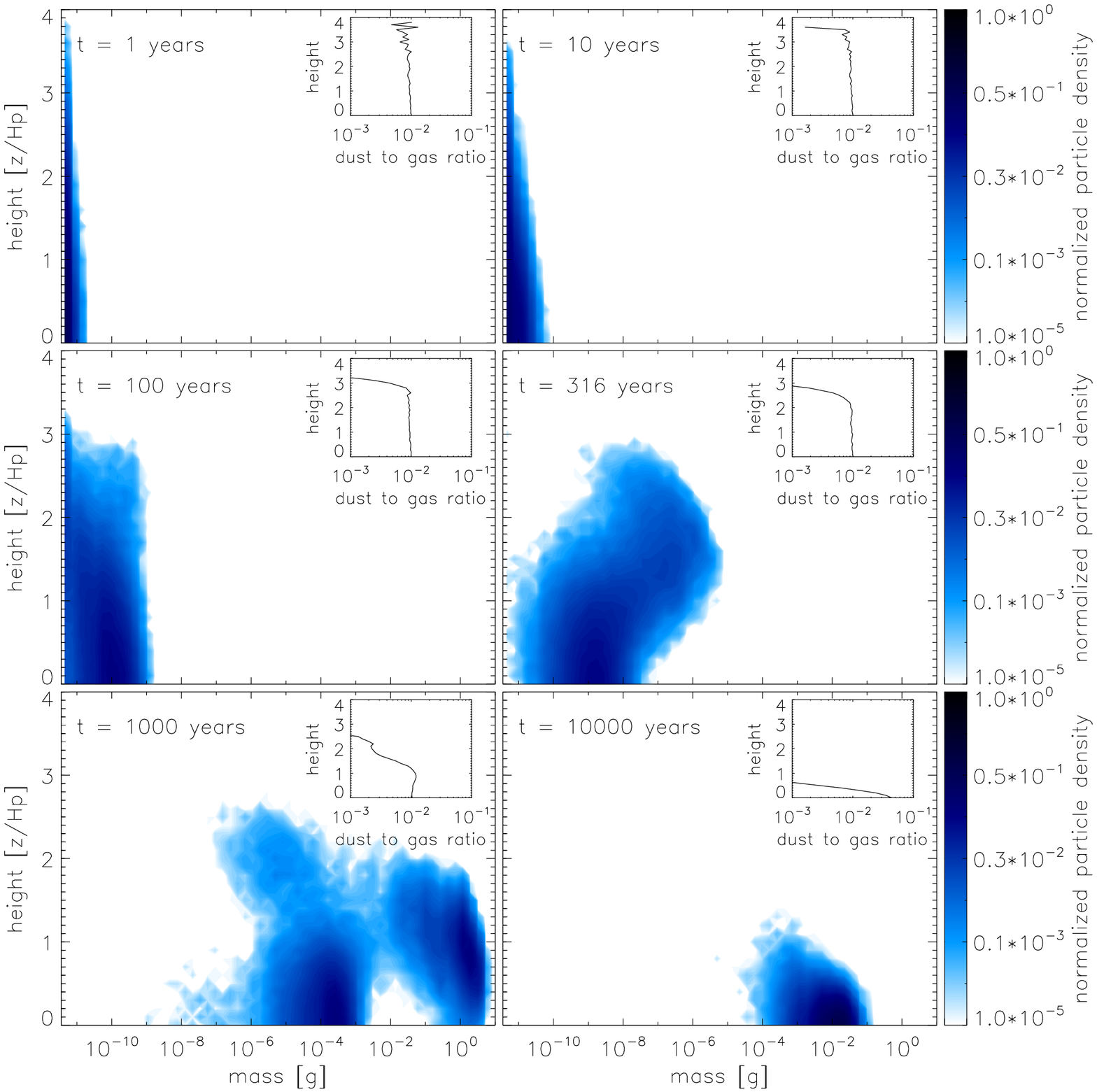}
  \caption{Mass distribution for the complete Braunschweig model using the Okuzumi porosity prescription with $\alpha=10^{-4}$ (CB1). The axes and contours are the same as in Fig. \ref{fig:mass_simpl_not_or}.}
  \label{fig:mass_br_1d-4_or}
\end{figure*}

\begin{figure*}
\centering
  \includegraphics[width=0.75\textwidth]{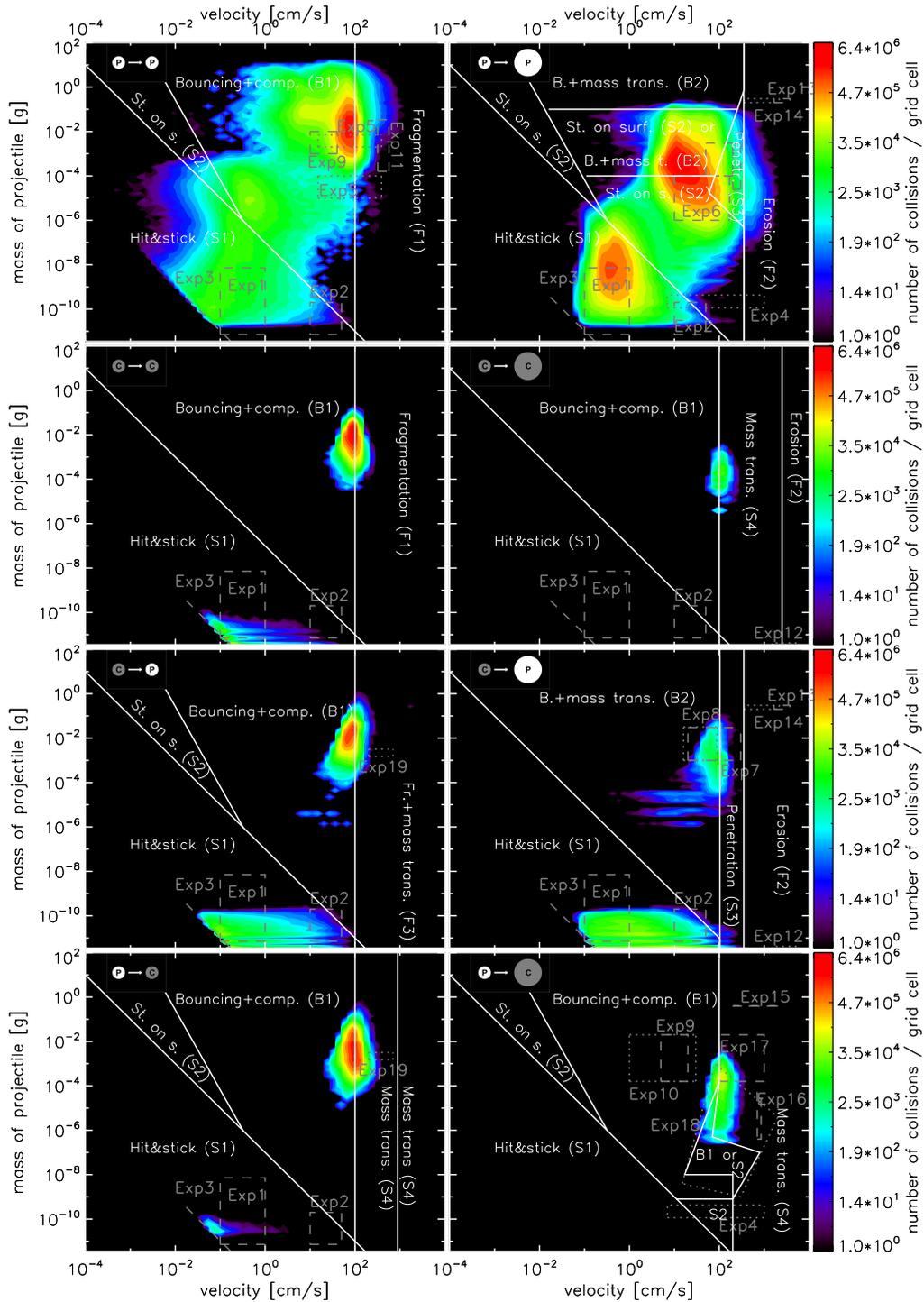}
  \caption{The collision history for the CB1 simulation. The eight regimes of the complete Braunschweig model are shown with the corresponding border lines of the nine different collision regimes (white solid lines). The x axis is the relative velocity of the particles in $cm/s$, the y axis is the mass of the projectile in gram units. The grey boxes indicate the areas that are covered with laboratory experiements (see Paper I for more details). The colors indicate how many collisions happened at the given part of the parameter space during the simulation. The red and yellow areas are `hot spots', where most of the collisions take place.}
 \label{fig:coll_hist_br_1d-4}
\end{figure*}

\begin{figure*}
\centering
  \includegraphics[width=0.75\textwidth]{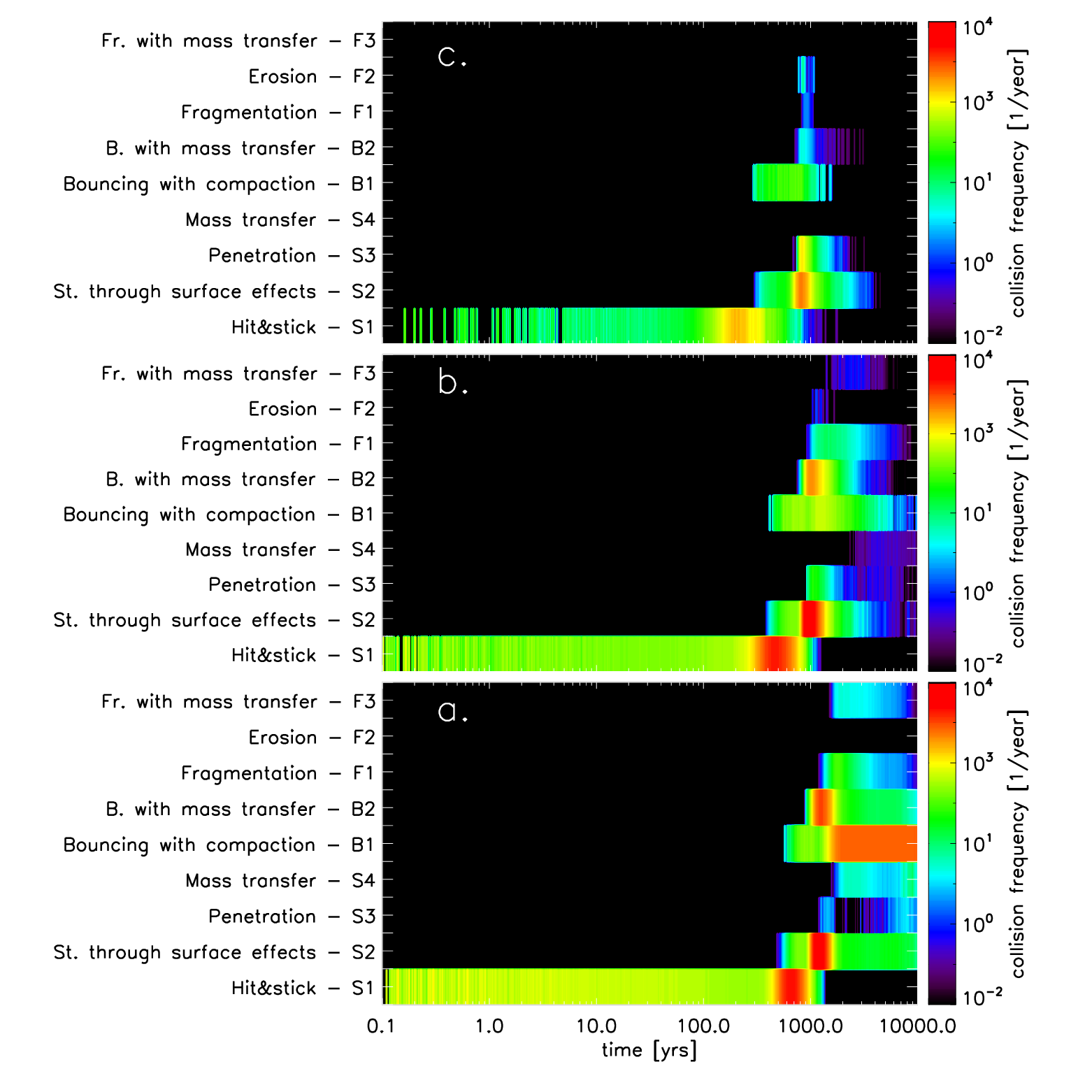}
  \caption{The collision frequency of the nine collision types for the CB1 simulation. The x axis is the time in years, the y axis indicates the collision types. The colors of the stripes indicate the collision frequency (e.g. the number of collisions per year). The collision frequency is shown at the midplane (a.), at 1 pressure scale height above the midplane (b.), and at 2 $H_g$ (c.).}
  \label{fig:coll_hist_br_1d-4_or}
\end{figure*}

\begin{figure*}
\centering
  \includegraphics[width=0.75\textwidth]{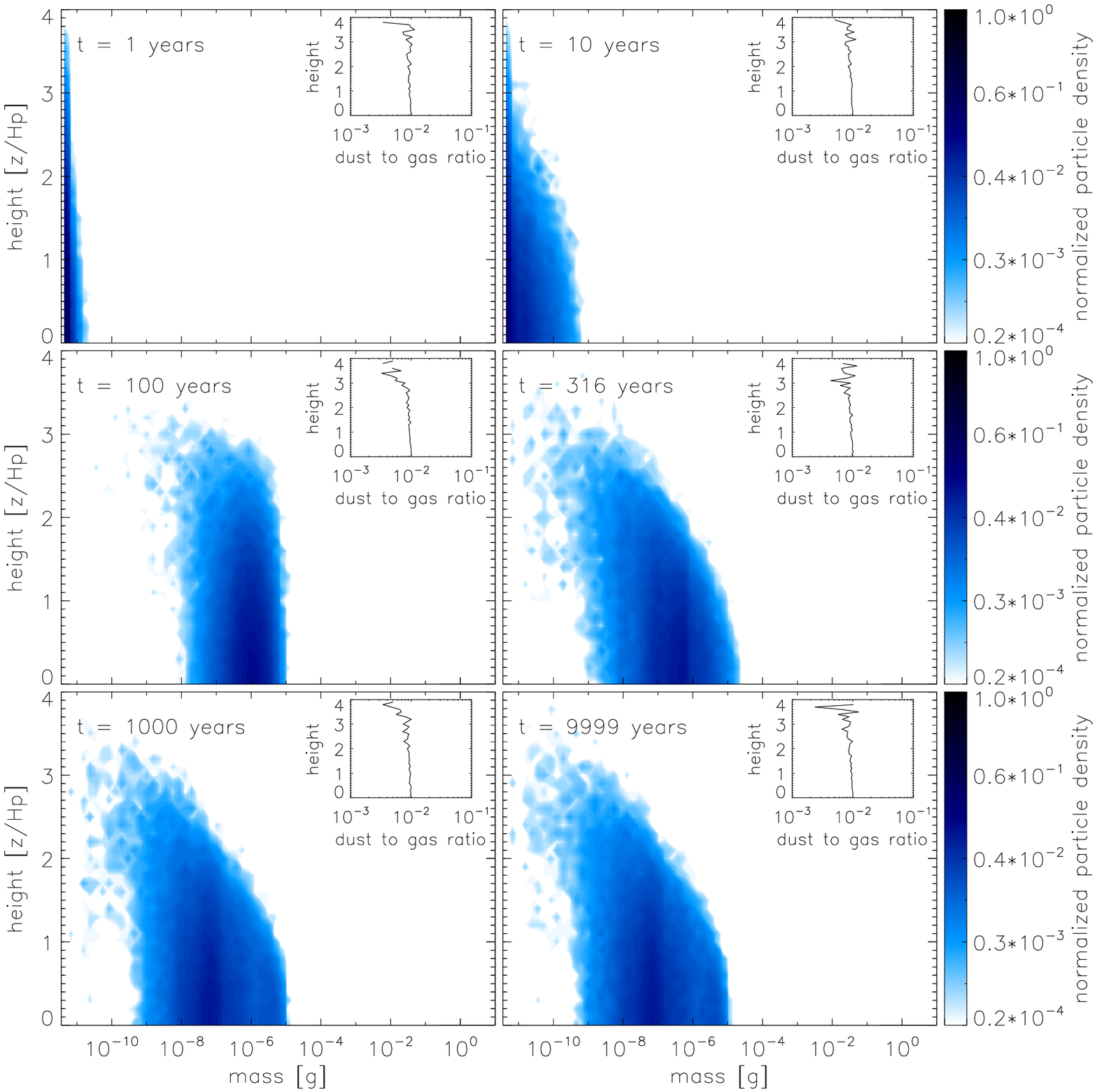}
  \caption{Mass distribution for the complete Braunschweig model using the Okuzumi porosity prescription with $\alpha=10^{-2}$ (CB3). The axes and contours are the same as in Fig. \ref{fig:mass_simpl_not_or}.}
  \label{fig:mass_br_1d-2_or}
\end{figure*}

\begin{figure*}
\centering
  \includegraphics[width=0.75\textwidth]{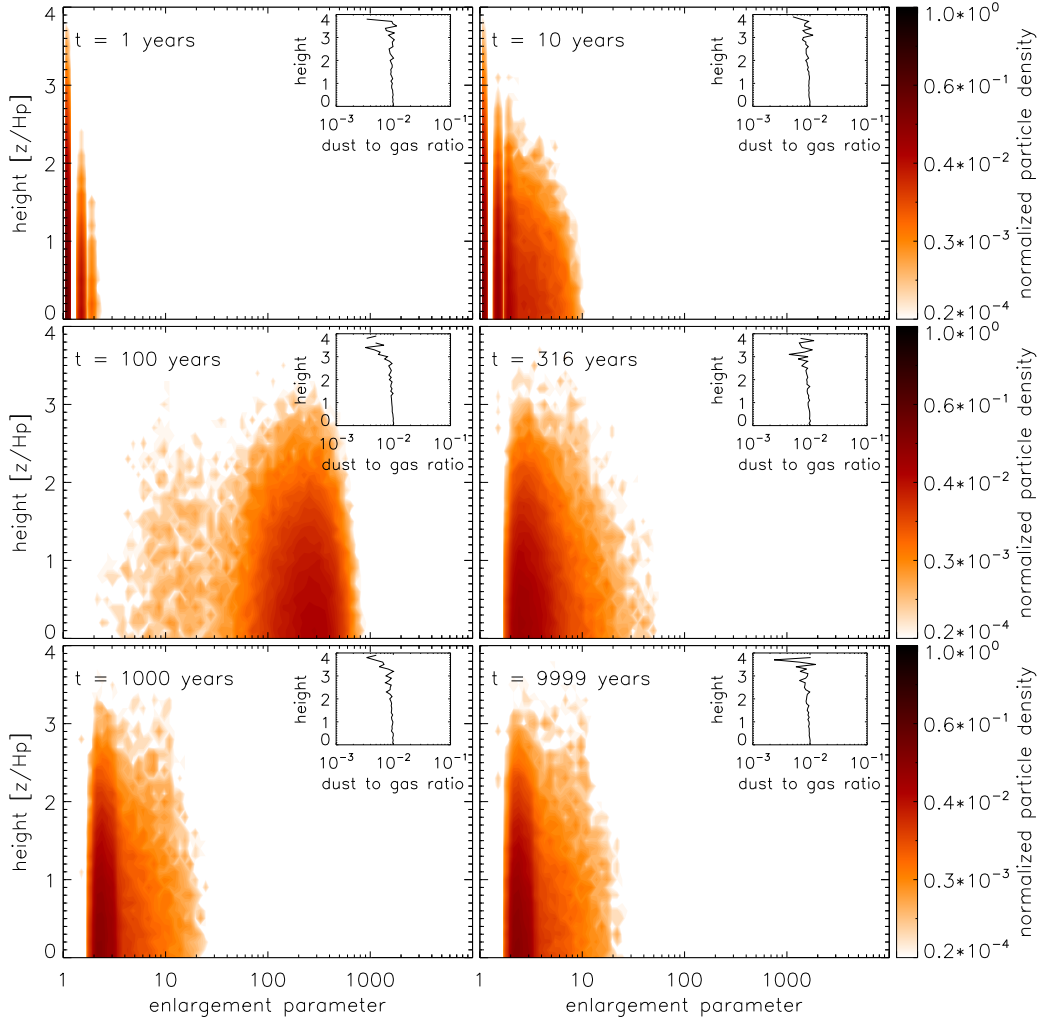}
  \caption{Enlargement parameter distribution for the complete Braunschweig model using the Okuzumi porosity prescription with $\alpha=10^{-2}$ (CB3). The x axis represents the enlargement parameter of the aggregates, and the contours show the normalized mass-weighted enlargement parameter of the particles.}
  \label{fig:enlargpar_br_1d-2_or}
\end{figure*}

\begin{figure*}
\centering
  \includegraphics[width=0.75\textwidth]{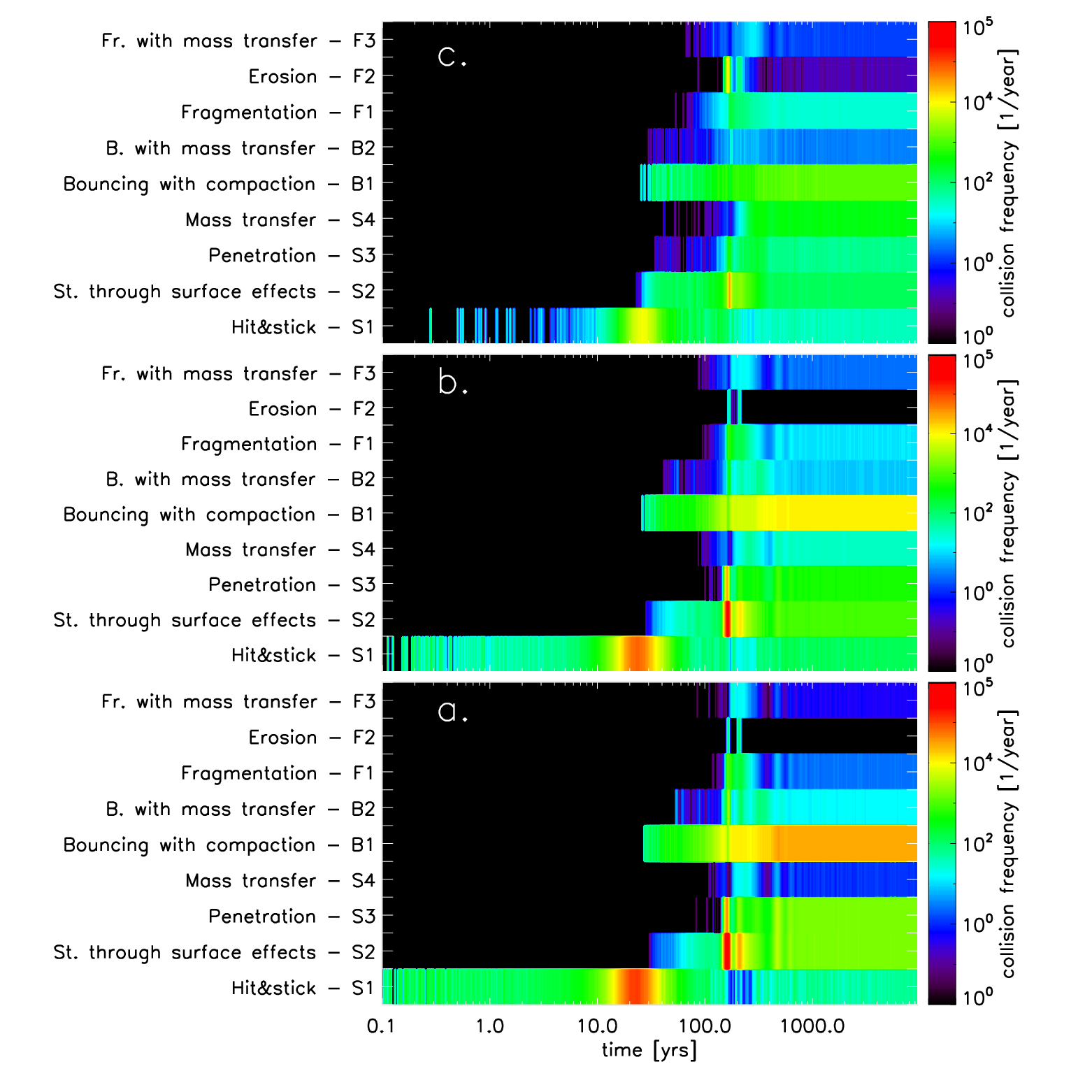}
  \caption{The collision frequency of the nine collision types for the CB3 simulation. The x axis is the time in years, the y axis indicates the collision types. The colors of the stripes indicate the collision frequency (e.g. the number of collisions per year). The collision frequency is shown at the midplane (a.), at 1 pressure scale height above the midplane (b.), and at 2 $H_g$ (c.).}
  \label{fig:coll_hist_br_1d-2_or}
\end{figure*}

\subsection{Porosity and collision models}
\label{sec:poro}
It is generally assumed that dust particles in the protoplanetary disk are aggregates, i.e. they consist of micron-sized spheres (monomers) which are connected via surface forces \cite{Henning1996}. The material of these monomers can be iron, silicate, organics (tar), different ices and a mixture of these (e.g., silicates with organic or icy mantels). The collision model described in Paper I is based on laboratory experiments performed on room-temperature dust aggregates made out of micron-sized silicate monomers. This naturally limits the region in the disk where our collision model is applicable. Therefore, our results are not applicable in regions where ices condensate (beyond the snow-line) or close to the central star where the effects of sintering (partial melting and crystallization) becomes important \citep{Poppe2003}. Another limitation that originates from the adopted collision model is that the laboratory experiments were performed using fractal dimension 3 aggregates. Therefore, we cannot reliably consider the initial fractal growth of the aggregates. This topic is further discussed in Sect. \ref{sec:collmod}.

The monomers are embedded in the gas and move relative to each other \cite{Beckwith2000}. The particles collide, stick, and grow due to their relative motions. There are two limiting cases for growth: cluster-cluster aggregation (CCA) and particle-cluster aggregation (PCA). CCA growth means that the aggregates tend to collide with other aggregates that consist of the same number of monomers. PCA growth occurs when an aggregate collides successively with much smaller aggregates (monomers) only. The resulting structure of the aggregates are different for the two growth types: CCA growth produces fluffy fractal aggregates with a fractal dimension of 2, while PCA growth produces compact structures with a fractal dimension of 3 \citep{Ossenkopf:1993p80}.

Two monomers can perform different types of motions while they are connected if the available kinetic energy is high enough. These are rolling, sliding, twisting, and the connection can break due to pull-off \citep{Dominik:1997p89}. Most important of these is rolling as the aggregate can dissipate the initial collisional energy efficiently through rolling motions. The rolling energy ($E_{\mathrm{roll}}$) is defined as the energy that is needed to roll two monomers by 90 degrees. During the initial stages of growth, the collisional energy is lower than the rolling energy, therefore no significant restructuring occurs. The aggregates stick at the first contact and we refer to this as the hit\&stick (S1) collision type.

We adopt the hit\&stick porosity model of \cite{Okuzumi2009a} in this paper. In Paper II we used the porosity model of \cite{Ormel:2007p93} which only treats PCA and CCA collisions and constructs semi-analytical recipes if the size (or mass) ratio of the two colliding aggregates are intermediate. However, \cite{Okuzumi2009a} improved on this by modeling these intermediate regions, so called quasi-CCA collisions (QCCA), where two clusters with a given mass ratio can collide. The porosity model of Ormel et al. produces aggregates with a fractal dimension of 2.5. The Okuzumi et al. model results in fluffy, fractal dimension 2 aggregates. As the Ormel model analytically interpolates between PCA and CCA collisions, whereas the Okuzumi et al. model directly simulates these collisions, we prefer the Okuzumi model as the fiducial hit\&stick porosity model.

As the particles grow, their collisional energy is increasing and we cannot neglect restructuring. This phase starts when the collisional energy is a few times larger than the rolling energy between the monomers. As compaction occurs, the fractal dimension of the aggregates increases. At the same time, the average number of connections a monomer has (coordination number) increases as well.

The further evolution of the dust aggregates is as of yet unclear and a matter of ongoing debate. Currently there are two competing collision models. The Braunschweig collision model (Paper I) is mostly based on laboratory experiments performed by using fractal dimension 3 aggregates (so called dust cakes and their pieces). As such, it postulates that the aggregates at one point during their evolution in the protoplanetary disk reach a rather compact state with a fractal dimension of 3. The validity of this assumption was not thoroughly checked yet (but see future work in Sect. \ref{sec:collmod}). This collision model predicts that nine different collision types can occur between dust aggregates of different sizes and porosities (e.g., bouncing, mass transfer, cratering, etc.).

The other alternative comes from molecular dynamics models. \cite{Wada2008} performed a series of collisions using equal sized aggregates of fractal dimension 2 with an ever increasing relative velocity. They found that the aggregates end up with fractal dimension 2.5 due to restructuring before fragmentation sets in. This would imply that the collision model can be constructed from hit\&stick collisions at low velocities, sticking with restructuring at higher velocities, and finally fragmentation. They did not observe any other collision type.


High relative velocities are required to initiate restructuring of equal sized collisions. Relative velocity due to Brownian motion is negligible for large aggregates, as $\Delta v_B$ decreases with mass. Relative velocity due to differential radial drift and settling is zero, as equal sized aggregates drift and settle with the same speed. Relative velocity due to turbulence is also zero or negligibly small, if both particles are well coupled to the gas \citep{Weidenschilling1993, Ormel:2007p92}. If one particle is decoupled from the gas, the relative velocity between different and equal sized aggregates is roughly constant. However, one needs sufficiently large aggregates to enter this regime.

The relative velocity between different sized aggregates is larger than between equal sized aggregates during the initial stages of growth. As a result, collisions between different sized collisions are more frequent. To support this statement, we use one simulation result obtained in Paper II: one of the Minimum Mass Solar Nebula (MMSN) models with ID Mt1d-4m100. We produced a logarithmically spaced mass grid array and we calculated how many collisions happened between the aggregates within each grid cell during the simulation. The number of collisions per grid cell is indicated by the contour levels in Fig. \ref{fig:coll_rate}. We only used the first 300 yr of the simulation because particles at $t>300$ yr experience other collision types than sticking. Since the relative velocity of small particles is dominated by Brownian motion, similar sized aggregates collide (CCA-like growth) as long as the particle mass is less than $10^{-5}$ g. However, when the turbulent relative velocity dominates over Brownian motion (at masses above $10^{-5}$ g), the collision rate between equal sized aggregates is reduced. The large aggregates (mass of $10^{-4}$ g) preferentially collide with smaller ones (mass of $10^{-8}$ g). Therefore the results of \cite{Wada2008} (which is based on equal sized collisions) might not capture the full complexity of the aggregate restructuring phase.

Ideally, a restructuring collision model considering different sized aggregates is needed. As such a model -- and its experimental verification -- is not available yet, we for simplicity decide to stick with the Braunschweig lab-based model, as we did in previous works. We acknowledge the potential caveats of our adopted collision model, which we further discuss and quantify in Sect. \ref{sec:collmod}.

\section{Prelude: the effects of porosity and collision models}
\label{sec:intro}
We perform altogether 21 simulations to investigate the effects of different porosity models and turbulence, in which we gradually use more realistic collision models. The IDs and parameters of these simulations are shown in Tab. \ref{table:sedi}. First we compare our model against the results of DD05 for compact particles (model DD in Tab. \ref{table:sedi}). Then we use the hit\&stick porosity model of \cite{Okuzumi2009a} to investigate the effects of porosity (model DDa). So far we assume that the aggregates stick at all relative velocities and that the turbulence parameter $\alpha$ is zero. In the next step we construct a more realistic collision model with sticking, bouncing and fragmentation (model SB1). We call this collision model the ``simplified Braunschweig model'' because it uses only three collision types out of 9, which is described in Paper I (the complete Braunschweig model). In the next step we turn on turbulence (models SB2-3) to examine the effects of turbulent stirring. Finally, we use the complete Braunschweig model with turbulence (models CB1-4) and use different disk models (models LD1-2 and HD1-2).

%


\subsection{Test: comparison with the DD05 model}
\label{sec:test}
DD05 performed a simulation (S2 in their paper, DD in this paper), where the disk model is the same as the one described in Sect. \ref{subsec:inicond}: the particles are compact, upon collision particles stick together at all collision energies, and the only source of relative velocity is Brownian motion and differential settling. They found that the `rain-out' particles reach the midplane in 500 yr having attained sizes of a millimeter (10$^{-2}$ g in mass). 

We performed the exact same simulation to validate our code. We find that the `rain-out' particles in our simulation reach the midplane also in 500 yr and have masses of $10^{-2}$ g. Therefore we can conclude that our code works properly. We illustrate the mass evolution of particles as a function of their height at six different snapshots ($t=1$ yr, 10 yr, 100 yr, 316 yr, 10$^3$ yr, 10$^4$ yr) in Fig. \ref{fig:mass_DD05}. 

Brownian motion is essential in our simulations because growth by Brownian motion initializes the sedimentation-driven coagulation. The reason is that we have initially a \textit{mono-disperse} particle size-distribution (meaning that all particles have the same size and mass). The aerodynamical properties of these particles are identical, thus there is no relative velocity due to settling between the monomers at a given height. If growth due to Brownian motion was not initiated (e.g. growth by Brownian motion did not introduce aggregates with different aerodynamical properties than that of the monomers), the monomers would simply sediment to the midplane without any growth. Although DD05 included Brownian motion, this effect would not have been present as that simulation started with a (narrow, but not infinitely narrow) size distribution.

As shown in Fig. \ref{fig:mass_DD05}, growth by Brownian motion is faster at the midplane due to the higher gas and dust densities (at $t=1$ and 10 yr). Once particles in the upper layer also start to grow by Brownian motion, sedimentation-driven coagulation starts and particles at the upper layers grow much faster than the aggregates at the midplane (at $t=100$ yr) because the absolute value of the settling velocity increases with height (see Eq. \ref{eq:set}). The heaviest particles sweep up the smaller particles while they sediment and further increase their settling velocity, resulting in a rain-out at $t=500$ yr. Once the first rain-out particles reach the midplane, they could only grow by Brownian motion, because at the midplane, the settling velocity of any particle is zero (see Eq. \ref{eq:set}). But the relative velocity due to Brownian motion for such heavy aggregates is negligible. Therefore, the particles that have reached the midplane, do not increase in mass any longer. 

\subsection{The effects of the hit\&stick porosity model}
In the previous section we used compact particles. In this section, we consider dust aggregates which are built from (sub)micron-sized solid spheres, called monomers. Such monomers collide with each other and form fluffy structures due to hit-and-stick collisions (collisions that result in sticking upon the first contact). In this section we assume that all collisions result in sticking. We include growth by Brownian motion and settling only.

We use here the hit\&stick porosity model constructed by \cite{Okuzumi2009a}. As discussed in Sect. \ref{sec:poro}, this collision model defines a third type of aggregation next to the PCA and CCA collisions, that is QCCA (quasi cluster-cluster aggregation - collisions between clusters with a predefined mass ratio). This model is based on simulations of hit\&stick collisions without restructuring.

The mass and the porosity evolution are shown in Figs. \ref{fig:mass_DD05_ok} and \ref{fig:enpar_DD05_ok}. The most striking property of this simulation is the maximum mass and porosity of the particles. We end up with particles of $10^{10}$ g in mass having an enlargement parameter of almost $10^8$ (while the compact radius of such an aggregate is some meters, the enlarged radius is several kilometers!). The stopping time in the Epstein regime (Eq. \ref{eq:ts1}) is proportional to $m/A$. As the Okuzumi-model produces aggregates with a fractal dimension of $\sim 2$ (the mass scales with $a^2$, where $a$ is the particle radius), the stopping time only slightly increases with mass. Therefore, particles settle slowly and produce extremely fluffy structures. At one point, however, the aerodynamical cross section radius becomes larger than the mean free path of the gas, and the aggregate enters the Stokes regime (Eq. \ref{eq:ts2}). As the stopping time is now proportional to $ma/A$, the stopping time more strongly increases with mass. This transition from Epstein to Stokes regimes happens at t=900 yr for the particles located at 1.7 $H_g$ above the midplane. Once the transition happens for a given particle, it settles to the midplane in a matter of years due to the heavy mass of the aggregate. Therefore, when the size of the aggregate becomes comparable to the mean free path of the gas ($a=\lambda_{\mathrm{mfp}}$), we reach a natural upper size limit where rain-out in any model is expected. 




We emphasize that any collision model containing exclusively sticking is only valid, if no significant restructuring happens during the collisions (e.g. the collision energy is less than $5 E_{\mathrm{roll}}$, the rolling energy, which is the energy needed to roll two monomers by 90 deg). This condition is clearly not met at all times in our simulations, e.g. the rain-out particles can have collision velocities with the swept up particles as high as several 10 m/s in these simulations. Such collisions would result in catastrophic fragmentation. Therefore, the results presented in this section should be considered as toy models.

We conclude here that porosity alone is not sufficient to explain the observations. Therefore, the hit-and-stick assumption is insufficient to describe all settling stages, since we expect that the condition $E<5 E_{\mathrm{roll}}$ will be broken at some point. Clearly, a more realistic collision model that include compaction and fragmentation of dust aggregates, is then needed. Next, we will investigate the consequences of including these physical regimes, first by using a simplified prescription.

\subsection{A simplified Braunschweig model}
In this section we use the simplified version of the collision model described in Paper I. We assume sticking and the increase of the porosity, if the collision energy is smaller than $5 E_{\mathrm{roll}}$. Bouncing with compaction is used if the collision energy is greater than $5 E_{\mathrm{roll}}$, but the relative velocity of the two aggregates is less than 1 m/s. Fragmentation occurs if the relative velocity of two aggregates is greater than 1 m/s. The recipe for mass and porosity evolution for bouncing and fragmentation is taken from Paper I (our hit \& stick (S1), bouncing with compaction (B1), and fragmentation (F1) collision types). We still assume that the particles grow by Brownian motion and settling only (the effects of turbulence is discussed in the next section), and we use the porosity model of \cite{Okuzumi2009a} for the hit\&stick phase.

The evolution of the mass is shown in Fig. \ref{fig:mass_simpl_not_or}. The mass distributions at $t=1$, 10, 100 yr are identical to Fig. \ref{fig:mass_DD05_ok}. At $t=316$ and $1000$ yr we see the effects of bouncing at the intermediate energies. The rain-out particles cannot increase their mass further, when they suffer bouncing collisions. Therefore, their masses are $10^{-2}$ g when they arrive to the midplane and no larger aggregates are formed as opposed to the DDa model. As the rain-out particles are smaller, they settle slower, and reach the midplane only at $t=6000$ yr. The enlargement parameter at the end of the hit\&stick phase is $\sim 10^{4}$ and it decreases significantly only in the midplane to values between 10-$10^4$ as the rain-out particles collide and bounce with the aggregates in the midplane. The particles in this simulation are always in the Epstein regime, as bouncing collisions prevent growth to sizes above the mean free path of the gas.


We conclude that realistic collision models reduce the particle sizes in sedimentation-driven coagulation and thus reduce the settling time. However these collision models with the combined effects of porosity are still insufficient to explain the long-term presence of the 10 micron feature in disk SEDs. We now include turbulence into the simulation.

\section{Results}
\label{sec:res}

\subsection{The effects of turbulence}
\label{sec:turb}

So far all particles sooner or later ended up at the midplane because there was no effect that could counteract settling. In this section we examine the effects of a non-zero turbulence parameter, which can stir particles back up. 

A small turbulence parameter ($\alpha = 10^{-6}$) does not significantly affect the masses of the rain-out particles compared to the SB1 model of the previous section (see model SB2 in Tab. \ref{table:sedi}). As particles do not only settle but also diffuse downward (and upward) due to turbulence, the time the rain-out particles reach the midplane is somewhat shorter than for model SB1. In all previous simulations a dense dust layer formed at the midplane of the disk. However, even this low level of turbulence can prevent the formation of this layer and introduce a non-zero (although small) dust scale-height. The porosity of the aggregates are also similar to the results described for the SB1 simulation.

The influence of turbulence is more pronounced if $\alpha=10^{-4}$. The mass evolution of the aggregates is shown in Fig. \ref{fig:mass_simpl_1d-4_or}. The first rain-out particles reach the midplane already at $t=500$ yr due to downward diffusion, although these particles have lower masses than in model SB1 ($10^{-4}$ g -- therefore, in the absence of turbulence, these particles would reach the midplane later than the particles in SB1). The porosity of the aggregates is strongly influenced by the higher turbulence value and bouncing. Due to the increased turbulent relative velocities, particles start bouncing before they reach the midplane. The average enlargement parameter is $2 \times 10^3$ at the end of the hit\&stick phase. As the particles settle to the midplane and reach an equilibrium height, bouncing continues and gradually compactifies the aggregates. The enlargement parameter at $t=10^4$ yr is between 2 and 100. The dust distribution reaches a quasi-steady state at $t=10^{4}$ yr. 

We see that the particle mass is constant as a function of height at $t=10^4$ yr. As turbulence effectively mixes the particles, and as bouncing prevents further growth or fragmentation (the dust growth is halted), both the masses and porosities of the aggregates are similar at all heights where dust is present. We will see in the next section that this is only true if the turbulence parameter is rather modest. A value of $\alpha=10^{-2}$ results in height-dependent particle mass. 

We also see from these simulations that a higher turbulence value reduces the mass of particles and increases the dust scale height. If turbulence is strong enough, the dust scale height can be similar to the gas scale height and the disk atmosphere remains dusty at all times. However, such high turbulence value prevents any significant dust growth, which is not a fertile environment for planet formation (see next Section). 

\subsection{The complete Braunschweig collision model}
\label{sec:comp_br}
In this Section we use the complete Braunschweig model (see Paper I for details), the value of the turbulence parameter is $\alpha=10^{-4}$ and $10^{-2}$. The calculations are performed with both the Okuzumi porosity model (CB1, CB3) and the Ormel porosity model (CB2, CB4) because we want to investigate how sensitive the outcome of dust growth is to the used hit\&stick porosity model within the context of our model.

In the simplified Braunschweig collision model, the growth is halted by bouncing immediately if the particles enter the bouncing regime. However, in the complete Braunschweig model, there channels for growth beyond the hit\&stick border line (that is where $E_{\mathrm{coll}} > 5 E_{\mathrm{roll}}$). The most important area is in the {\pP} regime where a small porous projectile collides with a heavy porous target (see Fig.11 of Paper I). Due to these ``green'' areas at intermediate collision energies, particles in the CB1 and CB2 simulations grow to higher masses than in the SB3 simulation. As a consequence, the scale height of the dust is lower in these simulations, as heavier particles are more difficult to stir up by turbulence. We illustrate the mass distribution at $t=1$, 10, 100, 316, $10^3$, $10^4$ yr in Fig. \ref{fig:mass_br_1d-4_or} for the CB1 model. 

The particle evolution has two phases in these simulations. The first 1000 yr are identical for the SB3 and CB1/CB2 simulations, respectively (see also the first five snapshots of Figs. \ref{fig:mass_simpl_1d-4_or} and \ref{fig:mass_br_1d-4_or}). In this phase, particles start sedimenting, and the rain-out particles reach the midplane. The dust evolution in the SB3 model halts at this point as only bouncing collisions happen. However, during the second phase of the CB1 and CB2 simulations, particles can grow to higher masses because there are areas in the parameter space that is favorable for growth somewhat beyond the bouncing barrier (see Paper I and Paper II for a detailed explanation). Due to this growth, particles reach $10^{-2}$ g in mass for the CB1 and CB2 simulations.

The time evolution of the enlargement parameter is quite different in the two cases. Fluffy aggregates are produced with enlargement parameter $\Psi = 10^3$ when the Okuzumi porosity model is used. However, these aggregates are strongly compacted by bouncing and their final enlargement parameter is between 2 and 20. When the Ormel porosity model is considered, the enlargement parameter never exceeds 20 and by the end of the simulation the enlargement parameter is between 2 and 10. As we see, the enlargement parameter evolved through very different ways, but the final enlargement parameters are not so much different for the CB1 and CB2 simulations.

Figure \ref{fig:coll_hist_br_1d-4} illustrates the collision history of the CB1 simulation. If we compare this figure to Figs. 5, 7 and 10 of Paper II, we see that the features are more smeared out in Fig. \ref{fig:coll_hist_br_1d-4} than in the other figures. As we simulate here several boxes at different heights above the midplane, the physical conditions (e.g. gas density and sedimentation velocity) at the midplane and at the upper scale heights of the disk are different, which is responsible for the smeared out features of Fig. \ref{fig:coll_hist_br_1d-4}. 

We investigate the collision frequency of the nine collision types as a function of time (see Fig. \ref{fig:coll_hist_br_1d-4_or}). If one compares this figure with Figs. 4c, 6c, and 9c of Paper II, we immediately see that the diversity of occurring collision types is much greater in the CB1 simulation, although the strength of the turbulence is the same in all cases ($\alpha=10^{-4}$). This can be explained by the presence of sedimentation. The particle population in a given box is not fixed as in Paper II. During the rain-out process, ``heavy'' particles coming from $z=1$-2 $H_g$ `hit' the particle population at the midplane. Due to this process, the relative velocity between the small, midplane particle population and the generally larger rain-out population is increased. Thus collision types can occur that require larger collision energies.

We explore the effects of strong turbulence (CB3 and CB4 runs). In general we find that a turbulence parameter of $\alpha = 10^{-2}$ is able to keep the upper layers of the disk atmosphere dusty (see Fig. \ref{fig:mass_br_1d-2_or}). However the size of the particles at the midplane is several order of magnitude smaller than in the CB1 or CB2 simulations. This is not favorable for planetesimal formation via self-gravity assisted planetesimal formation \citep{Johansen:2007p65,Cuzzi2008, Youdin2011}. The high level of turbulence lowers the enlargement parameter in the CB3 simulation (see Fig. \ref{fig:enlargpar_br_1d-2_or}). The value of the enlargement parameter does not exceed $10^3$ during the simulation. The final enlargement parameter is between 2 and 10 for both the CB3 and 4 simulations.

Interestingly, the particle masses as function of height are constant in the simulations with $\alpha=10^{-4}$ or $10^{-6}$, but for higher $\alpha$ values, this is not the case. The particles in the upper layers are significantly smaller than particles at the midplane. The particles are mass-sorted (i.e. heavy particles are mostly located at the midplane and small particles at the upper layers). As 100 micron-sized particles from the midplane is mixed upwards by the strong turbulence, it looses mass along the way due to fragmenting collisions. Similarly, as a particle moves closer to the midplane, it is growing in mass due to sticking collisions. This can be explained by the decreasing gas density as a function of height. At low densities the stopping time (and therefore the relative velocity) of a particle is high. At high densities the particles are stronger coupled to the gas, their stopping time decreases. Therefore the relative velocity between these particles are low and they can coagulate. Figure \ref{fig:coll_hist_br_1d-2_or} illustrates this effect as fragmentation (F1, F2 and F3) happens frequently at z=2 $H_g$ above the midplane (green shade for F1 collisions) but it is rare at the midplane (dark blue shade for F1 collisions). 

We can also verify this behavior by comparing the viscous and collision timescales in this model. The viscous timescale is 
\begin{equation}
t_v = \frac{H_g^2}{\nu_T},
\end{equation}
using the parameters of the disk model we get that $t_v=22$ yr. The collision timescale can be calculated as 
\begin{equation}
t_c = \frac{1}{\Delta v \sigma n},
\end{equation}
where $\Delta v$ is the relative velocity of the two particles (in this case it is 1 m/s, the critical velocity for fragmentation), $\sigma$ is the cross section of the two particles (in this case two monomers with 1 micron size as such particles are present at 4 $H_g$), $n$ is the number density of the particles (calculated at the height of 4 $H_g$ assuming monomer-sized particles). Using these values, we get that $t_c=8$ yr. Therefore it is further verified that particles are fragmented while they are mixed upwards by turbulence as the timescale for collisions is shorter than the timescale for mixing.

It is also important to note that the particles at the upper layers of the disk have masses of $10^{-10}$ g and below (size of $\sim$ 1-10 micron). Although the upper layers of the disk are not well resolved, the obtained particle sizes are comparable to the sizes required to explain the 10 micron feature of SEDs \citep{boekel2003} and this vertical dust distribution has reached steady state after $t=2000$ yr. Therefore, we conclude that in the framework of our models, high values of turbulence (therefore energetic collisions with strong bouncing and fragmentation) are needed to explain why the disk atmospheres are dusty throughout the lifetime of the disk.

\subsection{Different disk models}
We performed simulations in two additional disk models to explore the effects of the gas surface density. These are the low density (with $\Sigma_g = 10$ g/cm$^2$) and high density simulations (with $\Sigma_g = 1000$ g/cm$^2$). See also Tab. \ref{table:sedi}, simulations LD1-2 and HD1-2. The dust to gas ratio, the gas scale height and the gas temperature are the same as in all previous simulations, we change the turbulence parameter only. 

We find that the final mass and Stokes number of the particles depend on the gas density (see Tab. \ref{table:sedi}). The higher gas density increases the particle masses but decreases the Stokes numbers. Thus the low density model produces the smallest particles (10$^{-4}$ g if $\alpha=10^{-4}$) but these particles have the highest Stokes number amongst the three gas densities (almost $10^{-2}$ if $\alpha=10^{-4}$). This can be explained by considering the stopping time of these particles. The stopping time is inversely proportional to the gas density (see Eq. \ref{eq:ts1}). Therefore particles in the low density model have greater stopping times than particles in the high density model. Particles with large stopping times are loosely coupled to the gas and have therefore high collision velocities. For this reason, the maximum particle mass proportional to the gas density.

The common feature of these simulations is that the dust scale height for $\alpha=10^{-2}$ is always similar to the gas scale height (see Table \ref{table:sedi}). Therefore, we conclude that the disk atmospheres can be kept dusty in sparse and dense disks alike with sufficiently high turbulence values.

\section{Discussion and future work}
\label{sec:disc}

\subsection{Validity of the porosity model}
\label{sec:collmod}
The Braunschweig collision model is based on laboratory experiments that used fractal dimension 3 aggregates with an enlargement parameter typically between 7 and 2. As seen in Table \ref{table:sedi}, the maximum enlargement parameter of the aggregates can be several orders of magnitude larger than the dust aggregates used in the laboratory. More specifically, the hit\&stick collision regime is followed by a collision type called sticking through surface effects (see Paper I) and bouncing at higher collisional energies. Thus we assume that the transition from fractal aggregates in the hit\&stick regime to fractal dimension 3 aggregates in bouncing is an instantaneous one. Furthermore, our porosity model for bouncing is calibrated for dust cakes (enlargement parameter of 6), however the enlargement parameter at the end of the hit\&stick phase is 2-3 orders of magnitude larger than the enlargement parameter of the dust cakes (see Table \ref{table:sedi}). Therefore, it is questionable whether the porosity model for the restructuring phase is correct. Significant changes in the porosity of aggregates has the potential to significantly alter our collision model and thus the obtained results. 



It is still debated how the porosity and the fractal dimension evolves during the evolution of dust aggregates. As this is a central issue in determining the dust properties, we need detailed information on the porosity evolution of dust aggregates. To resolve these issues, we plan to follow the hit\&stick and restructuring phases of a particle distribution in a self-consistent way by combining the Monte Carlo model of ZsD08 with a molecular dynamics code in a follow-up paper. 

\subsection{Aggregates beyond the snow line}
\label{sec:ice}
The particle sizes at the midplane of the disk are rather small ($\sim$ 100 microns if $\alpha=10^{-2}$) as we see in the previous section. The question arises: under what conditions could planetesimals form via successive collisions of dust aggregates? Or how do large enough dust aggregates form that could, under favorable conditions, be concentrated in e.g. vortices or turbulent eddies, become self-gravitating, and form eventually planetesimals \citep{Johansen:2007p65, Cuzzi2008, Youdin2011, Johansen2011}?

One answer to these questions could be icy aggregates. The molecular dynamics simulations of \cite{Wada2008, Suyama2008} showed that icy aggregates are very resilient to restructuring. They observed sticking between icy aggregates at a relative velocity as high as 50 m/s. The uncertainty in these simulations is the microphysical parameters of the icy monomers (such as critical displacement, surface energy, Young modulus etc.). Recently laboratory experiments were performed using micron-sized ice monomers by \cite{Gundlach2011}. They measured the rolling energy between icy monomers and it turned out the previously assumed values by \cite{Wada2008, Suyama2008} are in good agreement with the measured laboratory values. If experiments also confirm that icy aggregates can stick at relative velocities as high as 50 m/s, that would provide a way to form large enough dust aggregates beyond the snow-line in the solar nebula.


\subsection{Is $\alpha$ constant as a function of height?}
\label{sec:alpha}
\cite{Gammie1996} proposed the concept of layered accretion disks. If the ionization fraction of the gas is not sufficient to support magneto-rotational instability (MRI - \cite{Balbus1991}), the turbulence parameter drops and a dead-zone forms at the midplane of the disk. The extent of the dead-zone is uncertain, as the ionization processes of the gas are not well-constrained. For typical TTauri disks it can extend between 0.1 - 4 AU \citep{D'Alessio1998}. Inside 0.1 AU, the thermal radiation from the star can keep the gas sufficiently ionized for MRI, and outside 4 AU, the gas surface density is typically below 100 g/cm$^2$, therefore cosmic rays can penetrate the disk and keep it MRI active at all heights. 

It was proposed by \cite{Okuzumi2009} that negative charges on the surfaces of grains could prohibit growth. As we use a Monte Carlo code to follow the evolution of aggregates, it is possible to include a third particle property: the amount of charges present on the grain. In order to follow the charge evolution of the grains, we need to solve for the ionization state of the gas (i.e. amount of charges available in the gas phase), how efficiently these charges are collected by the dust grains and how the charges affect the relative velocity of the aggregates. 

We plan to investigate how dust evolves in a layered disk model. Small dust particles can very efficiently sweep up charges in the gas. As shown by \cite{Turner2010}, the dead-zone can extend to 2 $H_g$ for 1 micron-sized particles, but it shrinks below 0.5 $H_g$ for aggregates that are 100 micron in size. In a simulation like the one presented here, this would mean that as the particles grow, the dead-zone shrinks. When the dead-zone disappears, the whole disk becomes MRI active and the particles settled to the midplane might be fragmented and stirred back up. This could lead to initial oscillations before an equilibrium state is reached. 




\section{Summary}
\label{sec:concl}
We performed simulations in a 1D vertical column of a protoplanetary disk to better understand the process of sedimentation. We simultaneously solved for the particle motion and growth inside this column. The complexity of the models was gradually increased to examine the effects of different processes. The first simulation used a collision model that only contained sticking. We furthermore assumed that the particles were compact, and the turbulence parameter ($\alpha$) was set to zero. Later on we investigated the effects of different porosity models, more realistic collision models (with sticking, bouncing and fragmentation) and turbulence of different strengths. Below we summarize our results.

\begin{itemize}
\item Porosity helps to produce heavier particles by decreasing the stopping time of the aggregates, but porosity alone cannot prevent rain-out.


\item If the size of the particle becomes greater than the mean free path of the gas, the drag law changes from Epstein to Stokes drag. At this point, rain-out becomes unavoidable due to the change in the drag law.

\item Realistic collision models with bouncing and fragmentation limit the maximum particle sizes to be not more than 1 mm - 1 cm (particle masses between $10^{-3}$ - 1 g ). The exact value depends strongly on the strength of the turbulence, and the gas density.

\item A higher value of turbulence decreases the particle masses but increases the dust scale height. Using the simplified Braunschweig model with $\alpha=10^{-6}$ results in a dust scale height of 0.2 $H_g$ and final particle mass of $10^{-3}$ g, however the dust scale height is $0.6 H_g$, and the final particle mass is $10^{-4}$ g when using $\alpha=10^{-4}$. 

\item The final particle size and Stokes number depends on the gas density. The mass of the particles is decreasing with decreasing gas density, however the Stokes number remains roughly constant.

\item When using the most detailed collision model (the complete Braunschweig collision model), we obtain particle masses of $10^{-2}$ g (with an average radius of 1 mm, and an average Stokes number of $4 \times 10^{-3}$) and a dust scale height of 0.2 $H_g$ upon using $\alpha=10^{-4}$. However, the dust scale-height is almost 1 $H_g$ and the final particle mass at the midplane is $10^{-7}$ g (with an average radius of 100 micron, and an average Stokes number of $10^{-4}$) upon using $\alpha=10^{-2}$. Therefore, a sufficiently high turbulence value can keep the disk atmosphere dusty but the absence of significant dust growth is not favorable for planet formation.

\item The maximum enlargement parameter of the aggregates during their evolution can be as high as $10^4$. However, our adopted porosity model for bouncing is based on laboratory experiments performed using enlargement parameter $\sim$ 6 dust cakes. Therefore the largest uncertainty of our model is the porosity evolution.

\item The dust particle mass as a function of height is not constant if the turbulence parameter is $\alpha = 10^{-2}$. We see that the particle mass/size is a decreasing function of the height. Particles are 100 microns in size at the midplane and a few microns at 4 pressure scale heights.

\item The micron-sized particles present in the upper layers are comparable to the sizes needed to explain the 10 micron feature of disk SEDs. Therefore in the framework of our model, high values of turbulence are needed to explain why disk atmospheres are dusty for $\sim 10^{6}$ yr.

\end{itemize}

\begin{acknowledgement}

A. Zsom acknowledges the support of the IMPRS for Astronomy \& Cosmic Physics at the University of Heidelberg. C.W.O. acknowledges the financial support from the Alexander von Humboldt Foundation. We thank our referee (Hidekazu Tanaka) for his comments that greatly improved the quality and clarity of the paper. We thank J\"urgen Blum and Carsten G\"uttler for the fruitful discussions during the project.

\end{acknowledgement}

\bibliographystyle{bibtex/aa}

\end{document}